\documentclass[preprint2,longabstract]{aastex}

\shorttitle{New Z-Machine for the 45-m Telescope}
\shortauthors{Nakajima et al.}

\begin{document}

\title{A New 100-GHz Band Two-Beam Sideband-Separating SIS Receiver for Z-Machine on the NRO 45-m Radio Telescope}

\author{T. Nakajima\altaffilmark{1}}
\affil{Nobeyama Radio Observatory, National Astronomical Observatory of Japan,\\
462-2 Nobeyama, Minamimaki, Minamisaku, Nagano 384-1305, Japan}
\email{nakajima@nro.nao.ac.jp}

\author{K. Kimura and A. Nishimura}
\affil{Department of Physical Science, Graduate School of Science, Osaka Prefecture University,\\
1-1 Gakuen-cho, Naka-ku, Sakai, Osaka 599-8531, Japan}

\author{H. Iwashita and C. Miyazawa}
\affil{Nobeyama Radio Observatory, National Astronomical Observatory of Japan,\\
462-2 Nobeyama, Minamimaki, Minamisaku, Nagano 384-1305, Japan}

\author{T. Sakai}
\affil{Institute of Astronomy, The University of Tokyo\\
2-21-1 Osawa, Mitaka, Tokyo 181-0015, Japan}

\author{D. Iono}
\affil{Nobeyama Radio Observatory, National Astronomical Observatory of Japan,\\
462-2 Nobeyama, Minamimaki, Minamisaku, Nagano 384-1305, Japan}

\author{K. Kohno}
\affil{Institute of Astronomy, The University of Tokyo\\
2-21-1 Osawa, Mitaka, Tokyo 181-0015, Japan}

\author{R. Kawabe and N. Kuno}
\affil{Nobeyama Radio Observatory, National Astronomical Observatory of Japan,\\
462-2 Nobeyama, Minamimaki, Minamisaku, Nagano 384-1305, Japan}

\author{H. Ogawa}
\affil{Department of Physical Science, Graduate School of Science, Osaka Prefecture University,\\
1-1 Gakuen-cho, Naka-ku, Sakai, Osaka 599-8531, Japan}

\author{S. Asayama}
\affil{Chile Observatory, National Astronomical Observatory of Japan,\\
2-21-1 Osawa, Mitaka, Tokyo 181-8588, Japan}

\and

\author{T. Tamura and T. Noguchi}
\affil{Advanced Technology Center, National Astronomical Observatory of Japan,\\
2-21-1 Osawa, Mitaka, Tokyo 181-8588, Japan}

\altaffiltext{1}{Present address : Solar-Terrestrial Environment Laboratory, Nagoya University, Furo-cho, Chikusa-ku, Nagoya, Aichi 464-8601, Japan, nakajima@stelab.nagoya-u.ac.jp}

\begin{abstract}
We have developed a two-beam waveguide-type dual-polarization sideband-separating SIS receiver system in the 100-GHz band for {\it z}-machine on the 45-m radio telescope at the Nobeyama Radio Observatory. The receiver is intended for astronomical use in searching for highly redshifted spectral lines from galaxies of unknown redshift. This receiver has two beams, which have 45$^{\prime\prime}$ of beam separation and allow for observation with the switch in the on-on position. The receiver of each beam is composed of an ortho-mode transducer and two sideband-separating SIS mixers, which are both based on a waveguide technique, and the receiver has four intermediate frequency bands of 4.0--8.0 GHz. Over the radio frequency range of 80--116 GHz, the single-sideband receiver noise temperature is lower than about 50 K, and the image rejection ratios are greater than 10 dB in most of the same frequency range. The new receiver system has been installed in the telescope, and we successfully observed a $^{12}$CO ({\it J}=3--2) emission line toward a cloverleaf quasar at {\it z} = 2.56, which validates the performance of the receiver system. The SSB noise temperature of the system, including the atmosphere, is typically 150--300 K at a radio frequency of 97 GHz. We have begun blind search of high-{\it J} CO toward high-{\it z} submillimeter galaxies.
\end{abstract}


\section{Introduction}

Over the last 15 years or so, the advent of large-format focal plane arrays of millimeter and submillimeter bolometer detectors has opened the window for the viewing of distant star-forming galaxies. For instance, bolometer arrays including the submillimeter common-user bolometer array (SCUBA; Holland et al.\ 1999) on the James Clerk Maxwell Telescope (JCMT), a bolometric camera (Bolocam; Glenn et al.\ 1998) on the Caltech Submillimeter Observatory (CSO), the Max-Planck millimeter bolometer (MANBO; Kreysa et al.\ 1998) on the IRAM 30-m telescope, the large APEX bolometer camera (LABOCA; Siringo et al.\ 2009) on the Atacama pathfinder experiment (APEX), AzTEC (Wilson et al.\ 2008) on the JCMT and the Atacama submillimeter telescope experiment (ASTE), and the 960-element bolometric receiver (Shirokoff et al.\ 2009) on the South Pole Telescope (SPT) have been used to conduct deep millimeter and submillimeter-wave surveys toward blank and biased fields (e.g., Borys et al. 2003; Coppin et al. 2006; Pope et al. 2006, Greve et al. 2004, 2008; Laurent et al. 2005; Bertoldi et al. 2007; Scott et al. 2008; Perera et al.2008; Austermann et al. 2010; Tamura et al. 2009; Wei$\beta$ et al. 2009b; Scott et al. 2010; Vieira et al. 2010; Aretxaga et al. 2011; Hatsukade et al. 2011), yielding $>$ thousand detections of millimeter/submillimeter selected galaxies (SMGs; e.g., Blain et al. 2002).  Some of the uncovered SMGs, such as radio-identified SMGs (e.g., Ivison et al. 2007) were followed up by means of optical spectroscopy, revealing that these SMGs are located at {\it z} = 2--3 (e.g., Chapman et al.\ 2003; 2005). However, about 70 \% of the bright SMGs tend to be dark in the optical wavelength because it is buried in dust (Iono et al. 2006, Younger et al.\ 2007; 2009, Hatsukade et al.\ 2010, Tamura et al.\ 2010), and it is difficult to perform spectroscopic observation in the optical and near-infrared wavelength. Therefore, our plan was to determine the redshift distribution of SMGs using a millimeter wavelength that is not susceptible to dust extinction using a {\it z}-machine. In general, {\it z}-machine is used to observations of high-{\it z} objects because it is composed of a wide bandwidth detector and spectrometer. These systems are the key instruments used in the accurate determination of the redshifts of these SMGs. Given the fact that any of the lines in the CO ladder can appear at a particular frequency, the measurement of a single line is not sufficient to determine the redshift of an object, so at least two lines must be observed in order to make an unambiguous estimate.

To date, several concepts for a ``{\it z}-machine'' have been developed and/or employed for different wavelength regimes. For example, the frequency range of the Zpectrometer (Harris et al.\ 2007) on the Green Bank Telescope (GBT) is 28.5--34.5 GHz, and this has been used to successfully observe CO ({\it J} = 1--0) toward four SMGs (Harris et al.\ 2010). The Redshift Search Receiver (RSR; Erickson et al.\ 2007) is a 74--110 GHz detector and spectrometer for the Large Millimeter Telescope (LMT), but in the early stage of its use it has been installed on the Haystack 37-m telescope, as well as on the Five College Radio Astronomy Observatory (FCRAO) 14-m telescope. Chung et al. (2009) have already observed CO ({\it J} = 1--0) toward 29 ultraluminous infrared galaxies (ULIRGs) using the RSR on the FCRAO 14-m telescope. Z-Spec (Glenn et al.\ 2007) is a 185--293 GHz band receiver on the CSO, which Bradford et al. (2009) have used to present the first broadband $\lambda$ = 1 mm spectrum toward the {\it z} = 2.56 Cloverleaf quasar. The redshift ({\it z}) and early universe spectrometer (ZEUS; Stacey et al.\ 2007) is an echelle grating spectrometer in the submillimeter band. Moreover, the eight mixer receivers (EMIR; Carter et al.\ 2012) on the IRAM 30-m telescope and Wei$\beta$ et al. (2009a) have been used to detect CO ({\it J} = 3--2) in the 3-mm band and CO ({\it J} = 5--4) in the 2-mm band from an optically/ultraviolet faint SMG. They were able to successfully determine the redshift of the SMG for the first time.

We have developed a new {\it z}-machine for the Nobeyama Radio Observatory (NRO) \footnote{The Nobeyama Radio Observatory is a branch of the National Astronomical Observatory of Japan, National Institutes of Natural Sciences. (http://www.nro.nao.ac.jp/index-e.html)} 45-m radio telescope in 100-GHz band. The detectors of this receiver are waveguide-type sideband-separating (2SB) mixers that can observe the upper sideband (USB) and lower sideband (LSB) simultaneously. The receiver has a wide radio frequency (RF) range and is suitable for the {\it z}-machine. Moreover, we use the heterodyne receiver, which has advantage of high frequency resolution compared with another {\it z}-machine. The backend systems are newly developed wide-band digital spectrometers, which have a highly flexible FX-type spectrometer that allows a maximum bandwidth of 32 GHz (2 GHz $\times$ 16 arrays) on 4,096 spectral channels (Kamazaki et al. 2011). The specifications of the backend system will be presented in a separate paper, and we will describe the frontend part of the {\it z}-machine of the 45-m telescope in this paper. The 2SB mixer is adopted for a part of the frequency band of the Atacama large millimeter submillimeter array (ALMA), which suggests that the 2SB receiver systems are more efficient than other SSB receivers (e.g., Nakajima et al. 2007; 2008). Over an RF range of 80--116 GHz, the SSB receiver noise temperatures of the mixers are lower than about 50 K. The image rejection ratios (IRRs) are greater than approximately 10 dB in most of the same range. Moreover, our {\it z}-machine has two beams, and the beam separation is approximately 45$^{\prime\prime}$; therefore, the two beams can be turned on in an alternate manner for observation using the position switch. This can significantly reduce the amount of dead time and facilitate the efficient observation of point sources.

Using the newly developed {\it z}-machine for the 45-m telescope, we observed a $^{12}$CO ({\it J}=3--2) emission line toward the cloverleaf as confirmation of the performance of the receiver system. This is the first astronomical observation that has been made using a two-beam 2SB SIS receiver system in the 100-GHz band. The SSB noise temperature of the system, including the atmosphere, is typically 150--300 K.

In the present paper, we describe the two-beam 2SB receiver system in the 100-GHz band for the 45-m telescope, and we demonstrate its performance. The results of test observation of a high-{\it z} object using the {\it z}-machine are also presented.

\section{Optics}
\subsection{Optical Design and on-on Observation}

The details of the antenna design of the 45-m telescope are described in Nakajima et al. (2008). Basically, the new receiverfs optics are similar in composition to those of previous receivers on the 45-m telescope, and we have designed the receiver optics so that they would be adapted to the antenna optics. We used the methods of both Gaussian optics (Goldsmith\ 1998) and physical optics to design the receiver optics, implementing such features as an ellipsoidal mirror and a corrugated feed horn. Figure 1 shows the design of the receiver optics. This receiver has two beams, which are hereafter referred to as Beam-1 and Beam-2. The radio-frequency (RF) signal is split to the right and to the left by a plane mirror shaped like triangular prism at the center, and is then focused by the ellipsoidal mirror upon each horn. Because there are two beams, it is possible to continuously observe the target source during the position switching: when the Beam-1 points the source, the Beam-2 observes the off position. Then, the Beam-1 observes the off position, whereas the Beam-2 targets the source again. In this way, we always observe the target object with either the Beam-1 or 2. Therefore, the receiver has no dead time, which facilitates more efficient observation than can be achieved with a single-beam receiver (figure 2). We call this observation method on-on observation.

First, the optics were designed using Gaussian optics, which are generally used for optical design in the millimeter wave-band. Assuming the edge taper level of the sub-reflector to be $-$12 dB, we designed the new optics based on Gaussian beam propagation (Chu\ 1983). The reflection angle at the plane mirror is 70$^\circ$, and the ellipsoidal mirror was designed with an edge taper level of $-$30 dB to ensure that the beam separation would be narrow. We fabricated the ellipsoidal mirrors using an NC milling cutter. The mirrors are fixed just above each horn by holders, each of which has a position- adjustment mechanism.

Next, we used physical optics techniques to evaluate the optical system using Gaussian optics. In this calculation, we used the GRASP9 analysis software package of the reflector antenna. Figure 3 shows the antenna beam pattern. The antenna directivity is 92.57 dBi, which is equivalent to an aperture efficiency $\eta$ of 0.81 for the 45-m telescope at 100 GHz. The maximum cross-polarization level and the first side lobe level are about $-$25 dB and $-$19 dB, respectively, relative to the main lobe level. The beam sizes of Beam-1 and -2 are 17.$\!''$5 and 19.$\!''$1, respectively, and the separation between them in the sky is approximately 45$''$. However, the actual performance such as $\eta$ may be somewhat worse by several percent because the calculation was made without blocking by the sub-reflector stays, surface error, or ohmic loss of the optical elements.

\subsection{Corrugated Horn}
Corrugated horns are commonly used with reflector antenna systems. Because corrugated horns can reduce the edge diffraction, improved pattern symmetry and reduced cross-polarization can be obtained (Clarricoats et al.\ 1984). Such improved performance was needed for a horn that is to be used over a wide range of the RF frequency. We repeatedly calculated the appropriate corrugation pattern from the basic design. The physical dimensions of the corrugated horn are shown in table 1. The definitions of the parameters of the table are described in figure 4 of Nakajima et al. (2008). Thus, we obtained a return loss lower than $-$23 dB, a maximum cross-polarization level lower than $-$20 dB, and good similarity between the calculated beam profiles of the E-plane and H-plane (figure 4). We fabricated the horn using the direct-dig method, rather than the electro-forming method. The direct-dig method employs a specially-shaped bit to directly dig the grooves using an NC lathe from the aperture of a horn (Kimura et al. 2008). The displacement of the horn position upon cooling at 4 K was measured by a laser displacement gauge. The results showed we measured that the horn moved 1.3 mm in the horizontal direction of the cold head of a refrigerator, and also that the height of the horn came down to 1.3 mm. Therefore, the position of the horn corrects this displacement, and was set up accordingly in the receiver dewar.

\section{Receiver}
\subsection{Receiver Configuration}

A block diagram of the receiver system is shown in figure 5, and a photograph of the receiver in the dewar is shown in figure 6. The vacuum window of the dewar is made of a Kapton (polyimide) film with a thickness of 50 $\mu$m and has a $\phi$ 45-mm aperture. The infrared (IR) shield is made of Zitex (G106) film with a thickness of 150 $\mu$m and has a $\phi$ 40-mm aperture that is set in the 40 K shield. The size of these apertures is thrice of the beam radius, which corresponds to an edge taper level of $-$30 dB.

This receiver is composed of an ortho-mode transducer (OMT) and two 2SB mixers, both of which are based on a waveguide technique, with an intermediate frequency (IF) quadrature hybrid in each beam. The two orthogonal polarizations are split using an OMT. The OMT consists of a square to smoothed double-ridge transition guide followed by a junction of two side arms with a central guide based on the double-ridged waveguide OMT (Asayama \& Kamikura\ 2009). The detail of the new OMT configuration will be presented in a separate paper (Asayama \& Nakajima\ 2012). The RF signal is down-converted to 4.0--8.0 GHz using a 2SB mixer, as described in detail in Asayama et al. (2004). The corrugated horn, the OMT, the two 2SB mixers, and the IF quadrature hybrid constitute a tower structure on the 4 K cooled stage. The size of the 4 K stage, which is made of oxygen free high conductivity copper and gold coating, is 250$\times$600 mm. The outputs of the 2SB mixer are the USB and the LSB, both of which are centered at 6 GHz. The signals are then amplified by a cryogenic low-noise high electron mobility transistor (HEMT) amplifier. Because these amplifiers are optimized for low noise, bandwidth, and gain, the use of 4--8 GHz cooled isolators prevents the reflection of signals between the IF quadrature hybrid and the HEMT amplifier. In total, eight isolators and eight HEMT amplifiers are fixed on the 4 K stage.

The SIS junctions adopted herein were developed at the Advanced Technology Center of National Astronomical Observatory of Japan. Their design is based on a proposed bow-tie waveguide probe with a parallel-connected twin junction (PCTJ; Noguchi et al.\ 1995). The main difference between the present mixer and conventional mixers is the use of a full-height to 1/5 reduced-height waveguide for waveguide-to-stripline transition of the SIS mixer. The linearly tapered waveguide impedance transformer was designed using a lumped-gap-source port provided by HFSS$^{\rm TM}$ (Asayama et al.\ 2003). The PCTJ was composed of Nb/AlO$_{x}$/Nb junctions, with each junction having an area of 5 $\mu$m$^{2}$; the normal state resistance of this device was approximately 20 $\Omega$.

The operation frequency range, which is 80--116 GHz, is limited by the frequency range of the local oscillator (LO) chain. The LO signals for each 2SB mixer of dual-polarization are independently generated by multiplying the output of the signal generators (SGs), which cover the range from 14.6--18 GHz, by 2$\times$3 multipliers. Therefore, the four IF signals from the two 2SB mixers can cover different regions of the RF frequency range in each beam. The LO waveguides, which are used outside of the dewar, are made of silver and silver coating; they were fabricated by Oshima Prototype Engineering Co. The insertion loss of the silver waveguide is smaller than that of a copper waveguide. For example, when a silver waveguide with a length of 100 mm is compared with a copper waveguide of the same length, then there is a smaller loss of about 0.2 dB with the silver waveguide.

The receiver employs a two-stage Gifford-MacMahon cryocooler, which has a cooling capacity of 1.5 W on the 4 K cold stage and has a power consumption of 7.5 kW (Sumitomo RDK-415D). The receiver dewar is evacuated to 10$^{-4}$ Torr before cooling. It takes 9.2 h to cool the mixer from room temperature to the operating temperature ($\sim$4 K).

\subsection{Receiver Noise Temperature}

Before installing the receiver system in the telescope, we evaluated its performance. The noise temperature of the 2SB receiver was measured in the laboratory using a standard Y-factor method, and using a room temperature load (300 K) and a liquid nitrogen load (77 K). A chopper wheel switched the input signal between the hot and cold loads. We calculated the receiver noise temperature (T$_{\rm rec}$) using the standard equation:
\begin{equation}
T_{\rm rec} =\frac{300 - 77 \times Y}{Y - 1},
\end{equation}
where
\begin{equation}
Y = \frac{P_{\rm hot}}{P_{\rm cold}}.
\end{equation}
The temperatures of the hot and cold loads were 300 K and 77 K, respectively. The power meter readings when the hot and cold loads were presented to the input of the receiver were P$_{\rm hot}$ and P$_{\rm cold}$. Figure 7 shows the typical double sideband (DSB) mode noise temperature of the eight DSB mixers and the typical 2SB mode noise temperature of the four 2SB mixers. The measured receiver noise temperatures of the mixers of both the DSB and 2SB mode by a power meter, and the output IF of the receiver, were filtered using a fixed-bandpass 4 GHz filter ({\it f}$_{c}$ = 6 GHz). The mixer was mounted on a 4 K stage in a dewar. The first-stage IF amplifier is a 4 K cooled HEMT in the 4.0--8.0 GHz band. The equivalent noise temperature and the gain of the HEMT amplifier associated with an isolator were approximately 8 K and +30 dB, respectively. The following-stage amplifiers work at room temperature. 

The DSB receiver noise temperatures included the noise contributions from the vacuum window, feed horn, and IF amplifier chain, and were approximately 25 K over the LO frequency range of 80--115 GHz, which corresponds to 4--5 {\it hf / k}. The measured DSB noise temperature, except for the noise temperature of the cooled isolator and the HEMT amplifier (8 K), is consistent with the simulated results. The SSB receiver noise temperatures of each sideband were measured to be lower than approximately 50 K over an RF frequency range of 85--120 GHz. The corrected noise temperature of the SSB mode was calculated as follows:
\begin{equation}
T_{\rm SSB} =T_{\rm rec}(1 + \frac{1}{IRR}),
\end{equation}
where IRR is the image rejection ratio. Here, we have adopted values of 13.6 dB for the LSB and 11.0 dB for the USB, respectively, which were used to measure the average value of this receiver (see the next subsection). A minimum value of $\sim$ 34 K in the USB was achieved at approximately 106 GHz. The mean value and the standard deviation between the RF frequency range of 80--110 GHz in the LSB and 86--120 GHz in the USB were 48 $\pm$ 6 K and 41 $\pm$ 5 K, respectively.

\subsection{Image Rejection Ratio}

Claude et al.\ (2000, ALMA MEMO 316) \footnote{ALMA MEMO No.316 $\langle$http://www.alma.nrao.edu/memos/html-memos/abstracts/abs316.html$\rangle$} calculated the effect of the total amplitude and phase imbalances on the IRRs of a 2SB mixer. The main factor in the degradation of amplitude and phase imbalance is the performance of the quadrature hybrid of the RF and IF. In our design, we have minimized the amplitude and phase imbalances of the RF quadrature hybrid between the two paths through the receiver (Asayama et al. 2004). The amplitude and phase imbalance of the RF quadrature hybrid were measured to be about $\pm$ 2.5 dB and $\pm$ 5 degrees, respectively, using a Millimeter-wave Vector Network Analyzer. The IF quadrature hybrid was fabricated by the Japan Communication Equipment Co., Ltd. and the amplitude and phase imbalance were less than $\pm$ 0.5 dB and $\pm$ 3 degrees, respectively. Therefore, we could expect to achieve an IRR of greater than 10 dB.

The IRRs were measured by the relative amplitudes of the IF responses in the USB and LSB during the injection of a Continuous Wave (CW) signal (Kerr et al.\ 2001, ALMA MEMO 357) \footnote{ALMA MEMO No.357$\langle$http://www.alma.nrao.edu/memos/html-memos/abstracts/abs357.html$\rangle$} from a SG. The IRR measured by 6.0 GHz IF was greater than 10 dB in most parts of the RF frequency range of 79--116 GHz, as shown in figure 8. The mean values between the RF frequency range of 79--104 GHz in the LSB and 86--116 GHz in the USB are 13.6 dB and 11.0 dB, respectively. This IRR was a typical value for one of the four 2SB mixers, but other 2SB mixers had about the same performance.

\subsection{Installation}

The receiver system was installed in the 45-m telescope in May 2009. Figure 9 shows the receiver dewar in the receiver cabin. The RF signal is fed from the upper side, and the LO oscillators and IF chains are placed around the dewar. The receiver was set on the x,y stage and was put on the center of the beam axis using a plumb. The pressure level of the dewar and the temperature of the 4 K stage were 8 $\times$ 10$^{-9}$ Torr and 4.2 K, respectively, on the telescope.

Following the installation, the position of the receiver was adjusted by comparing the observation of the SiO maser source R-Leo with that of the other receiver in the 40 GHz band, which is the receiver for pointing using the SiO maser source. The beam squint between these receivers is smaller than 0.6$''$ at 86 GHz, a value that is less than 1/20 of the half power beam width (HPBW).

\section{Results}
\subsection{System Noise Temperature}
We installed the receiver system in the telescope and measured the performance of the receiver system. Note that the noise temperatures in this subsection are not corrected for IRR. Even after correction for IRR, the noise temperature increases by only ${\lesssim}$ 10 \%. The SSB noise temperatures of the system, including the atmosphere, were approximately 150--300 K at {\it f}$_{\rm RF}$ = 97 GHz in the USB during the test observations for all 2SB mixers. The system noise temperature, including the atmosphere, became approximately half of that of the previous single-beam receiver system, S100. 

The noise temperature of the optics between the 1st plane mirror at the Cassegrain focus and the 5th plane mirror on the way to the receiver was measured by a standard Y-factor method to be approximately 20 K, and the noise temperature of the optics, including the main and sub reflectors, was measured to be about 50 K. In total, eight mirrors are used without a main or sub reflector in the optics of the 45-m telescope, and many of them are made from carbon-fiber-reinforced plastics (CFRP) coated with dotite, which is a conductive adhesive agent. The losses of these mirrors are higher than that of a metallic mirror. Therefore, if the material of the mirror is changed, the system noise temperature may be reduced.

\subsection{Image Rejection Ratio}

We have already developed and operated the measurement system of the IRR for the 2SB receiver on the telescope (Nakajima et al.\ 2010). This system is comprised of a horn that is moved by a motor slider, a harmonic mixer, and an SG; the accuracy of measurement of the IRR was $\pm$ 10 \%. We upgraded the measurement system to improve the accuracy of measurement. The multiplication of signals from the SG does not use the harmonic mixer, but rather, a 2$\times$3 multiplier, in order to prevent spurious radiation in the new measurement system. The horn is set in a position distant from the receiver in order to prevent the occurrence of a standing wave. As a result, the accuracy of the measurement of the IRR was measured to be $\pm$ 5 \%. 

Figure 10 shows the relationship between the IRR value measured by the IRR measurement system and that estimated from the molecular line observation toward the W51 molecular cloud. We measured the integrated intensities and IRR values as a function of bias voltages of the 2SB mixers, because IRR value can depend on the bias voltage. Each measurement was conducted continuously five times. The standard deviation is shown as an error bar of the IRR. The integrated intensity of the signal was measured by the observations of $^{13}$CO ({\it J} = 1--0) in the USB and that of the image was measured in the LSB. The error margin calculated from the rms noise level is shown as the vertical error bar. The bold line shows the expected value of IRR according to the following IRR calculation formulae, and the dashed lines show the error ranges of $\pm$ 10 \% and $\pm$ 5 \%. The integrated intensity of the signal band (W$_{\rm signal}$) and image band (W$_{\rm image}$) are 
\begin{equation}
W_{\rm signal} = \frac{W_{\rm real}}{1 + \frac{1}{IRR_{\rm USB}}}
\end{equation}
and
\begin{equation}
W_{\rm image} = \frac{W_{\rm real}}{1 + IRR_{\rm LSB}},
\end{equation},
respectively. It is assumed that the optical depth of the signal band and image band are equal. Where W$_{\rm real}$ is the strongest integrated intensity. We adopted 152.7 K km$^{-1}$ from the observation. The error between the IRR measured by the measurement system and that measured by the observation ranges is from $-$4.5 to +7.7 \% for the signal band (figure 10 (a)) and from $-$2.2 to +10.7 \%, with no points of poor quality for the image band (figure 10 (b)). At a minimum, the measured values of the IRRs are within an accuracy of $\pm$ 5 \% in the signal band within the margin of error. The error of the IRR is mainly due to the accuracy of the signal intensity reading by the spectrum analyzer. The large deviation from the calculation value in the image band is thought to be due to a low signal-to-noise ratio.

\subsection{Beam Size and Main-Beam Efficiency}

We estimated the beam size and the main-beam efficiency of the 45-m telescope based on observations of the quasar 3C273 and of Saturn in continuum. Note that the following data was obtained at the beginning of the commissioning of the new receiver system, and thus the results are preliminary. We scanned 3C273 in the azimuth and elevation directions, while recording the total power of the IF output. The beam pattern of the antenna was obtained from the resulting map of Saturn by rotating it 180$^\circ$ on the peak position. Figure 11 shows the resulting map of Saturn at 86 GHz. The HPBW of the telescope is estimated to be 18.$\!''$93 $\pm$ 0.04 for Beam-1 and 18.$\!''$90 $\pm$ 0.04 for Beam-2, assuming that the beam pattern has a Gaussian shape. The beam patterns of both beams are nearly circular and show no distortion, though the effect of the sub-reflector stays can be seen. The main-beam efficiency, $\eta_{\mathrm{mb}}$, was calculated to be 39.5 $\pm$ 2.5 \% and 41.0 $\pm$ 1.3 \% for Beam-1 and Beam-2, respectively, assuming that the brightness distribution of Saturn is a uniform disk with an apparent diameter of 17.$\!''$03 and a brightness temperature of 153 $\pm$ 5 K. The separation between Beam-1 and Beam-2 was measured to be 46.$\!''$3 $\pm$ 2, as obtained by alternate pointing observation of Beam-1 and Beam-2, using the SiO maser sources, R-Leo and Ori-KL. This result is almost consistent with the designed value of the receiver optics (see section 2).

\section{Test Observations}
\subsection{First light}
Following the installation, the first astronomical signal from the inside of our galaxy and from the external galaxy were obtained on the $^{12}$CO ({\it J}=1--0) spectra at 115.271 GHz from the IRC+10216 late-type star on May 23, 2009 and the $^{12}$CO ({\it J}=1--0) spectra from the Markarian 231 active galaxy on March 11, 2010, respectively. Moreover, we successfully observed the $^{12}$CO ({\it J}=3--2) emission line at 97.191 GHz toward a cloverleaf quasar at {\it z} = 2.56 on December 18, 2011 (figure 12). The total integration time was about 1 hour and typical SSB system noise temperature was 250 K. Thus, our purpose of this observation is the operation check of the total telescope system. However, this was the first astronomical observation obtained using the two-beam 2SB receiver system in the 100 GHz band.

\subsection{Frequency Coverage}
We observed the molecular lines toward the Orion KL region by using new sixteen digital spectrometers with bandwidths of 2 GHz and spectral resolution of 488.24 kHz. The total frequency coverage is 32 GHz and we can detect the data of 16 GHz frequency range in one observation with this receiver, because the observation frequency range of Beam-1 and Beam-2 are common. Figure 13 shows the results, which were obtained by a single pointing in Orion KL. This image demonstrates the wide band capability of this receiver. We successfully detected the wide frequency range spectrum, which cover from 80 to 116 GHz. Since four IF signals of different frequency setting for each beam can be observed independently and simultaneously. We can detect this 36 GHz coverage data with just three local oscillator settings, thanks to the dual (USB + LSB) IF sidebands, each covering 4--8 GHz.

\subsection{Noise Level of on-on Observation}
The achieved rms noise level as a function of the observing time was measured with test observations. The rms noise of two beam receiver (on-on observation) is expected to $\frac{1}{\sqrt[]{2}}$ compare with single beam receiver (on-off observation), because the integration time is doubled in the case of two beam receiver during the same observing time. Figure 14 shows the results of this observation. For single beam receiver, the rms noise is decreases as the square root of the integration time in theory. We confirmed the rms noise level of two beam receiver became $\frac{1}{\sqrt[]{2}}$ relative to single beam receiver in accordance with our expectation.

\section{Conclusions}

We developed a two-beam waveguide-type dual-polarization sideband-separating SIS receiver system with the 100 GHz band for {\it z}-machine on the 45-m radio telescope at the Nobeyama Radio Observatory, Japan. The receiver is intended for astronomical use in searching for highly redshifted spectral lines from galaxies of unknown redshift. 

This receiver has two beams, with 18.$\!''$93 $\pm$ 0.04 for Beam-1 and 18.$\!''$90 $\pm$ 0.04 for Beam-2 of the HPBWs. It has 46.$\!''$3 $\pm$ 2 of beam separation and allows on-on position switch observation. The receiver of each beam is composed of an OMT and two 2SB mixers, both of which are based on a waveguide technique, and has four intermediate frequency bands of 4.0--8.0 GHz. 

The SSB receiver noise temperatures of each sideband were measured to be lower than approximately 50 K over an RF frequency range of 85--120 GHz. A minimum value of $\sim$ 34 K in the USB was achieved at approximately 106 GHz. The mean value and the standard deviation between the RF frequency range of 80--110 GHz in the LSB and 86--120 GHz in the USB were 48 $\pm$ 6 K and 41 $\pm$ 5 K, respectively. The IRRs measured by 6.0 GHz IF were greater than about 10 dB in most of the RF frequency range of 79--116 GHz. The mean value between the RF frequency range of 79--104 GHz in the LSB and 86--116 GHz in the USB were 13.6 dB and 11.0 dB, respectively. 

The new receiver system was installed in the telescope, and we successfully observed a $^{12}$CO ({\it J}=3--2) emission line toward a cloverleaf quasar at {\it z} = 2.56 on December 18, 2011. We have also begun a search of high-{\it J} CO toward high-{\it z} submillimeter galaxies (Iono et al.\ 2012). Moreover, we confirmed the wide frequency observation capability, which is 36 GHz band width, of this receiver with observation in Orion KL. For two beam receiver, the rms noise level is divided by square root relative to the the single beam receiver in accordance with our expectation. The achieved rms noise level is decrease in proportion to the $\frac{1}{\sqrt[]{2}}$ of integration time in theory.

\acknowledgments

The authors would like to thank Kazuyuki Muraoka, Tomomi Shimoikura, Sachiko Onodera, Bunyo Hatsukade, Tomohisa Yonezu, Tetsuya Katase, Junki Kizawa, and the 45-m group members for their contributions to this work. We are also grateful to Akira Mori and the entire staff of the Nobeyama Radio Observatory for their useful discussions and support. This work was supported in part by a Grant-in-Aid for Specially Promoted Research (20001003) and by a Grant-in-Aid for Young Scientists (B) (22740126) from JSPS.

\clearpage

\begin{figure}
\plotone{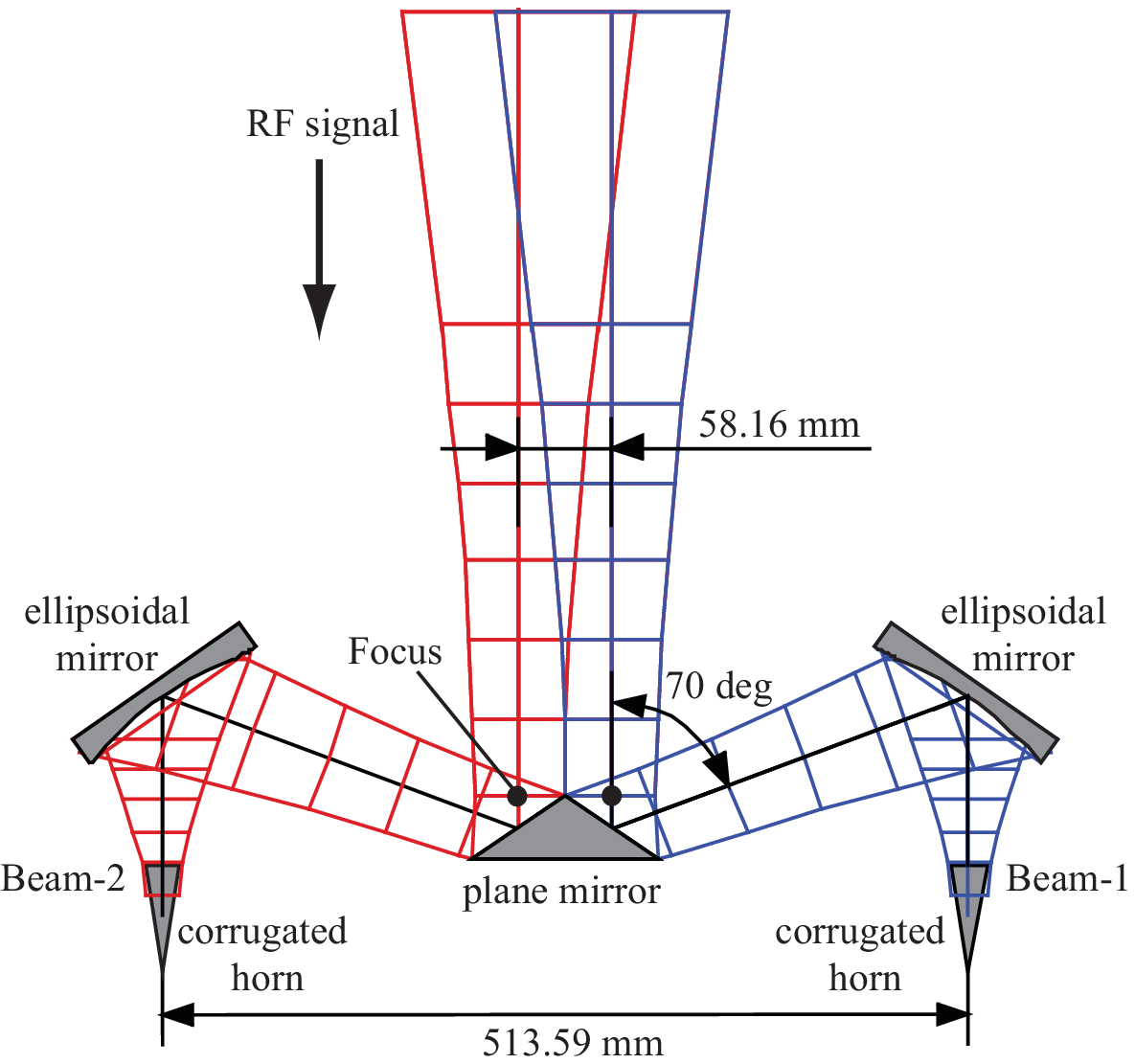}
\caption{Receiver optics of the new receiver. The RF signal is fed from the upper side. The separation of focus in each beam is 58.16 mm and the distance between corrugated horns is 513.59 mm.}\label{fig1}
\end{figure}

\begin{figure}
\plotone{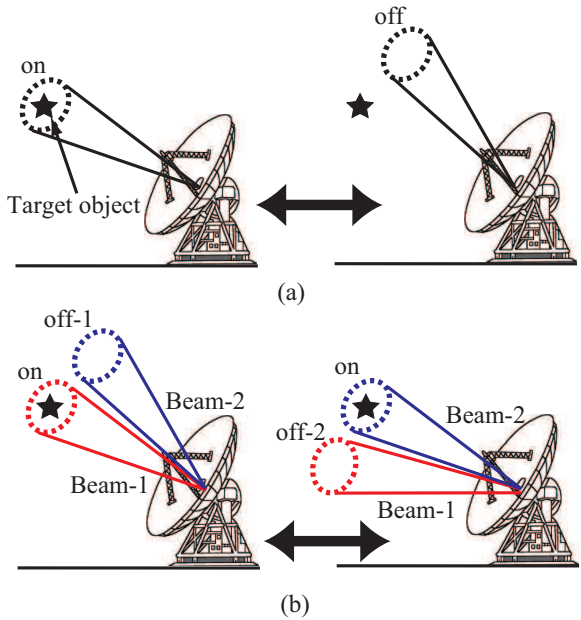}
\caption{The schematic images of the principle of position switch mode in comparison with two type receivers. The observation of on-off observation with single beam receiver (a) and on-on observation with two beam receiver (b).}\label{fig2}
\end{figure}

\begin{figure}
\plotone{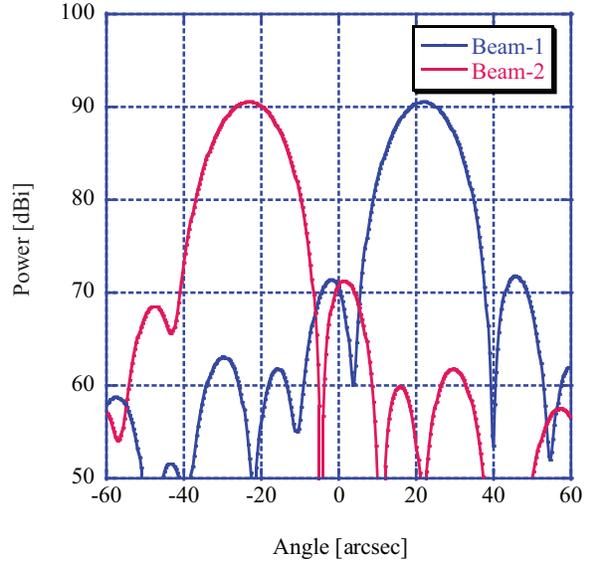}
\caption{Antenna beam patterns of Beam-1 and Beam-2 at 100 GHz, as calculated by GRASP9 physical optics software.}\label{fig3}
\end{figure}

\begin{figure}
\plotone{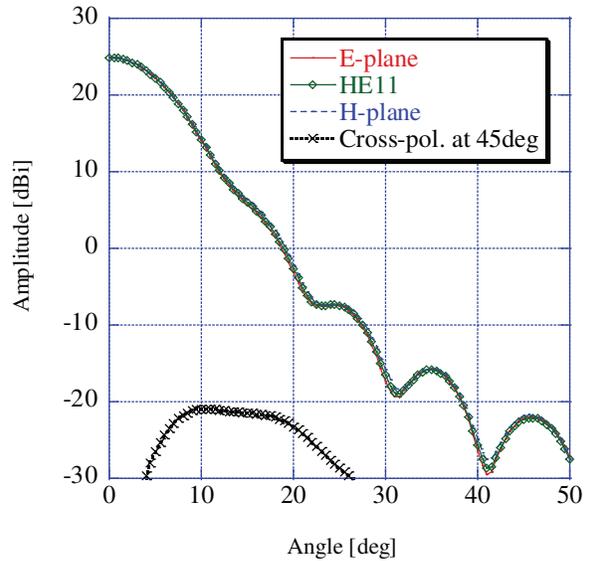}
\caption{Simulated beam profile of the corrugated horn at 100 GHz. The result shows the similarity between the beam patterns of the {\it E}-plane and the {\it H}-plane. These beam patterns correspond well with the theoretical beam pattern obtained by the radiation of the HE$_{11}$ mode.}\label{fig4}
\end{figure}

\begin{figure}
\plotone{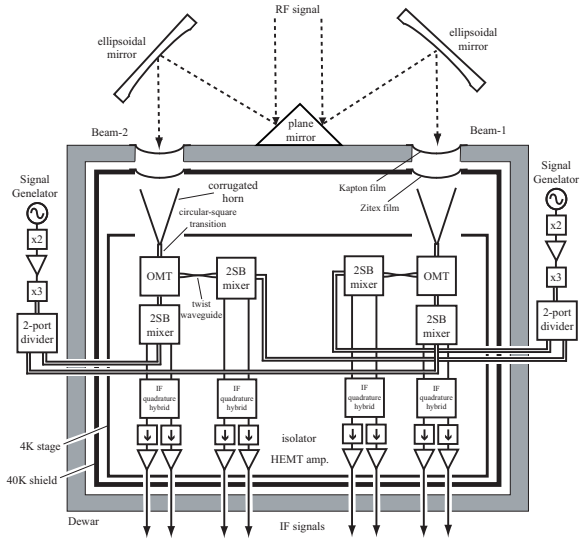}
\caption{Schematic block diagram of the receiver. The RF signal is fed from the upper side to each horn. The four IF signals are obtained independently and simultaneously.}\label{fig5}
\end{figure}

\begin{figure}
\plotone{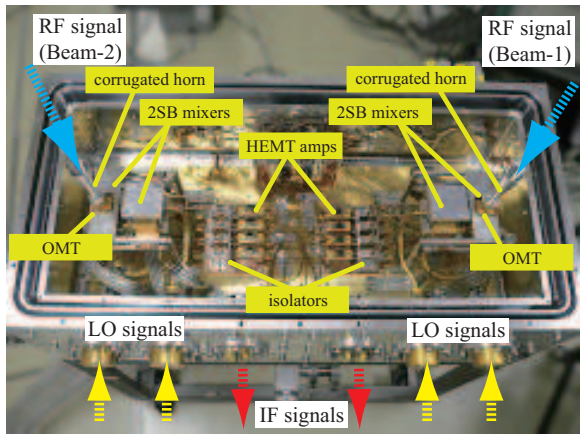}
\caption{Photograph of the 4 K cooled stage in the receiver dewar. The receiver components of Beam-1 and Beam-2 are symmetrically fixed on the 4 K stage. The corrugated horn, the OMT, two 2SB mixers, and two IF quadrature hybrids together constitute a tower structure.}\label{fig6}
\end{figure}

\begin{figure}
\plotone{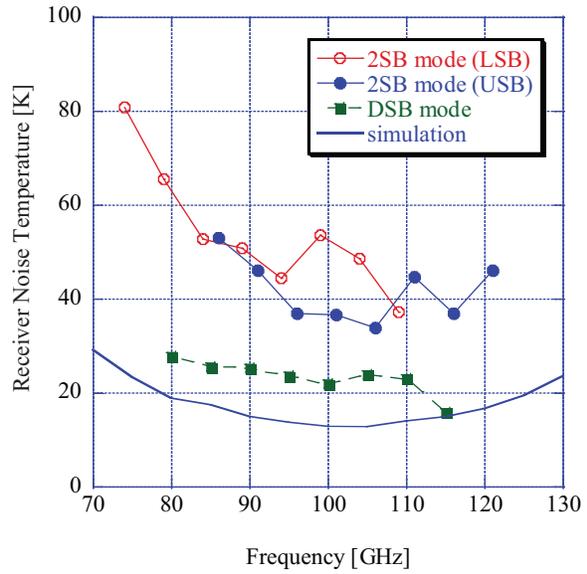}
\caption{The receiver noise temperatures of the DSB and 2SB (SSB) modes, and simulation of the noise temperature of the DSB mode. The square symbol represents the DSB mode and the solid line represents the simulation. The open and filled circle symbols represent the 2SB mode in LSB and USB, respectively. The horizontal axis is defined as an LO frequency for the DSB mode and an RF frequency for the 2SB mode, respectively.}\label{fig7}
\end{figure}

\begin{figure}
\plotone{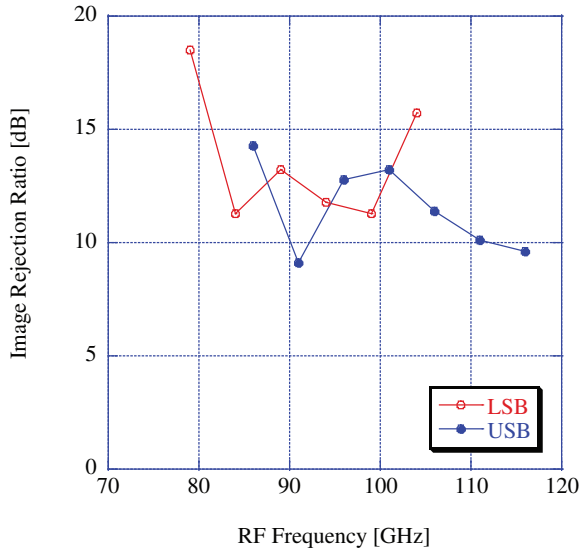}
\caption{Image rejection ratio in the LSB (open circles) and USB (filled circles). These values are measured by the 6.0 GHz IF and are shown as a function of the LO frequency.}\label{fig8}
\end{figure}

\begin{figure}
\plotone{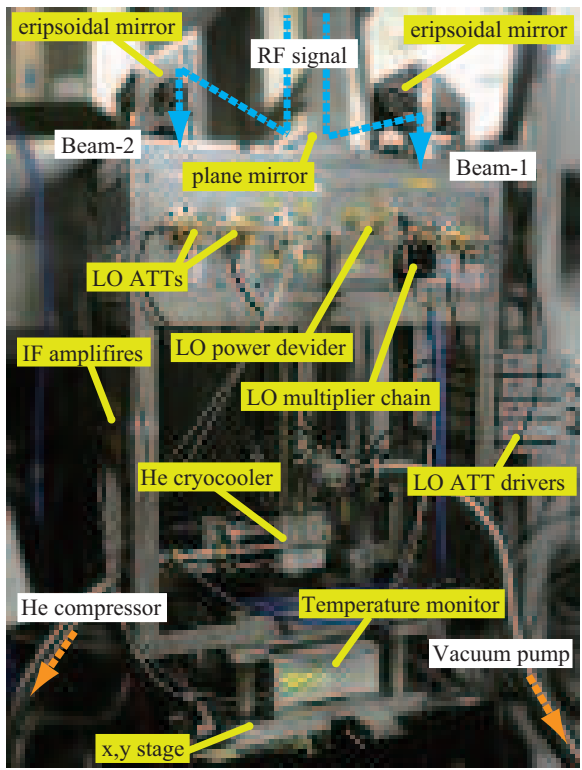}
\caption{Photograph of the receiver system in the receiver cabin of the telescope.}\label{fig9}
\end{figure}

\begin{figure}
\epsscale{.60}
\plotone{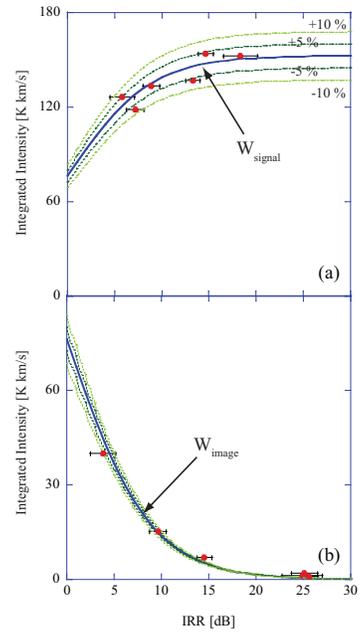}
\caption{The relationship between the IRR value as measured by the IRR measurement system and the integrated intensity as estimated from the molecular line observation in USB (a), and LSB (b), respectively. The bold line shows the anticipated value of the IRR according to the IRR calculation formula, and the dashed lines represent the error ranges of $\pm$ 10 \% and $\pm$ 5 \%.}\label{fig10}
\end{figure}

\begin{figure}
\plotone{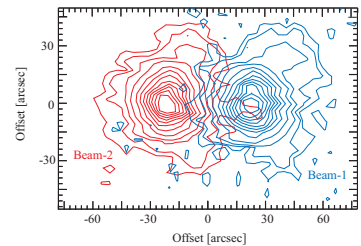}
\caption{Map of results for Saturn at 86 GHz. The lowest contour is 1 \% of the maximum value, and the intervals are 2 \%, 5 \%, and from 10 \% to 90 \%, at intervals of 10 \%. The elongated feature in three directions is the effect of the sub-reflector stays.}\label{fig11}
\end{figure}

\begin{figure}
\plotone{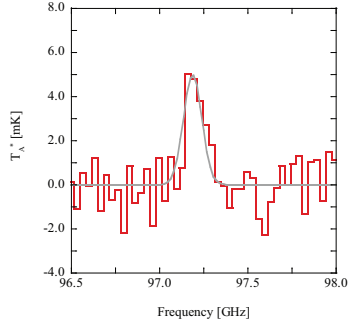}
\caption{Results of $^{12}$CO ({\it J}=3--2) emission line toward a cloverleaf quasar at high-$z$. The high-$J$ CO line, whose rest frequency is 345.796 GHz, is redshifted to the 100-GHz band.}\label{fig12}
\end{figure}

\begin{figure}
\plotone{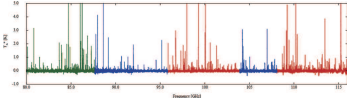}
\caption{Line spectrum of the Orion KL taken during commissioning of two beam receiver in May 2012. Almost 36 GHz of RF frequency space can be surveyed with just three local oscillator settings. The local frequencies of set-1 are 108 GHz for one polarization and 104 GHz for the other polarization (red), set-2 are 100 GHz and 96 GHz (blue), and set-3 are 92 GHz and 88 GHz (green).}\label{fig13}
\end{figure}

\begin{figure}
\plotone{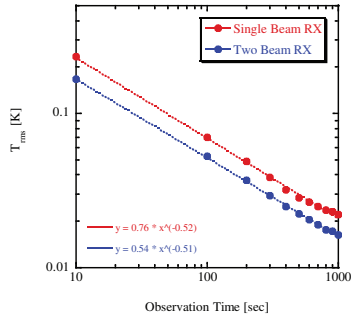}
\caption{The relationship between the rms noise level and observing time. The rms noise is decreases as the square root of the integration time in theory for single beam receiver (open circles) and for two beam receiver (filled circles). For two beam receiver, the rms noise level is divided by square root (dashed line) relative to the the single beam receiver (dotted line) in accordance with our expectation.}\label{fig14}
\end{figure}

\clearpage

\begin{table}
\begin{center}
\caption{Physical dimensions of the corrugated horn.}
\begin{tabular}{lc}
\tableline\tableline
parameters & value\\
\tableline
Material & Aluminum\\
Total length & 80.00 mm\\
Length of the corrugation section & 58.14 mm\\
Outer diameter of the aperture & 25.00 mm\\
Inner diameter of the aperture & 21.50 mm\\
Semi-flare angle & 9.0$^\circ$\\
Diameter of the circular waveguide & 3.08 mm\\
Width of the grooves & 0.57 mm\\
Depth of the grooves at the mode-launching section & 1.34--0.87 mm\\
Depth of the grooves at the thread section & 0.87 mm\\
\tableline
\end{tabular}
\end{center}
\end{table}

\end{document}